\documentclass{ws-procs10x7}

\begin{document}

\title{ Neutrino Masses Without Seesaw Mechanism \\
in a SUSY SU(5) Model With Additional $\overline{5}'_L+5'_L$}

\author{Yoshio Koide }

\address{Department of Physics, University of Shizuoka\\
52-1 Yada, Shizuoka, Japan 422-8526 
\\E-mail: koide@u-shizuoka-ken.ac.jp}

\twocolumn[\maketitle\abstract{
A radiatively-induced neutrino mass matrix with a simple 
structure is proposed on the basis of an SU(5) SUSY GUT model 
with $R$-parity violation.  The model has matter fields 
$\overline{5}'_{L}+5'_{L}$ in addition to the ordinary matter 
fields $\overline{5}_{L}+10_{L}$ and Higgs fields 
$H_u+\overline{H}_d$.  The $R$-parity violating terms are 
given by $\overline{5}_{L} \overline{5}_{L} 10_{L}$, 
while the Yukawa interactions are given by $\overline{H}_d 
\overline{5}'_{L} 10_{L}$.  Since the matter fields 
$\overline{5}'_L$ and $\overline{5}_L$ are different from 
each other at the unification scale, the $R$-parity violation 
effects at a low energy scale appear only through the 
$\overline{5}'_L \leftrightarrow \overline{5}_{L}$ mixings.
In order to make this $R$-parity violation effect harmless
for proton decay, a discrete symmetry Z$_3$ and a triplet-doublet 
splitting mechanism analogous to the Higgs sector are assumed.
}]

\section{Introduction}

Why do the neutrinos have such tiny masses?  
There are typical two ideas of the origin of neutrino masses:
One is the so-called ``seesaw  mechanism\cite{seesaw}, and 
the other one is  the 
``radiative mass-generation mechanism\cite{Zee}. 
The former can be embedded into a grand unification theory (GUT),
but the latter  is hard to be embedded into GUT.
For example, a supersymmetric (SUSY) model with $R$-parity violation 
can provide radiative neutrino masses\cite{R_SUSY}, but the model
inevitably induces unwelcome proton decay\cite{Smirnov}.
Therefore, as an origin of the neutrino masses, the
idea of the seesaw mechanism is currently influential
concerned with a GUT model.
However, the unified description of quark and lepton
mass matrices based on a GUT model is not still achieved 
even if we take the former standpoint.
In the present talk, 
against the current opinion,
I would like to investigate another possibility that 
the neutrino masses are radiatively generated.

The basic idea \cite{SUSY-Koide,SUSY-Koide2} is as follows: 
We introduce  matter fields $\bar{5}'_L +5'_L $
in addition to the matter and Higgs fields 
$\bar{5}_L +10_L + \bar{H}_d + H_u$ in the conventional minimal
 SUSY SU(5) GUT model.
The model has Yukawa interactions $\bar{H}_d\bar{5}_L 10_L$ 
and $R$-parity violation-terms $\bar{5}'_L \bar{5}'_L 10_L$.
Since the two $\bar{5}$-plet fields, $\bar{5}_L$ and 
$\bar{5}_L$, in the Yukawa interactions and $R$-parity 
violating terms, respectively, are different from each other, 
the $R$-parity violation-terms become visible only through
$ \bar{5}_L \leftrightarrow \bar{5}'_L$ mixing. 
In order to make the $R$-parity violation
harmless for proton decay, we will assume a mechanism 
analogous to a triplet-doublet splitting in the Higgs sector.

The explicit model is as follows:
We introduce a discrete symmetry Z$_3$ and assign the Z$_3$ quantum 
numbers as follows: 
\begin{equation}
\bar{H}_{d(-)} + {H}_{u(+)} + ({ \bar{5}_L} + 10_L)_{(+)} + 
( \bar{5}'_{L} + {5}'_{L})_{(0)} ,
\end{equation}
where $(+,0,-)$ denote the Z$_3$ transformation
properties $(\omega^{+1}, \omega^{0}, \omega^{-1})$ \ 
($\omega=e^{i 2\pi/3}$).
The Z$_3$ invariant tri-linear terms are only three:
\begin{eqnarray}
W_{tri} & =&  (Y_u)_{ij} H_{u(+)} 10_{L(+)i} 10_{L(+)j} \nonumber \\[4pt]
 & &{} + (Y_d)_{ij} \bar{H}_{d(-)} \bar{5}'_{L(0)i} 10_{L(+)j} 
 \nonumber \\[4pt]
 & &{}  + \lambda_{ijk} \bar{5}_{L(+)i} \bar{5}_{L(+)j} 10_{L(+)k} .
\end{eqnarray}
Note that $\bar{5}'_L$ in the Yukawa interactions are 
different from $\bar{5}_L$ in the $R$-parity violation-terms.
On the other hand, the Z$_3$ invariant bi-linear terms are only two.
In order to give ``triplet-doublet splitting",
we assume the following ``effective" bi-linear terms:
\begin{eqnarray}
W_{bi}& =& \bar{H}_{d(-)} (\mu + g_H \langle \Phi_{(0)} \rangle) 
H_{u(+)} \nonumber \\[4pt]
 &&{} +  \bar{5'}_{L(0)i} (M_5 - g_5 \langle \Phi_{(0)} \rangle) 
5'_{L(0)i}  ,
\end{eqnarray}
where $\Phi$ is a 24-plet Higgs field and its vacuum expectation
value (VEV) is
$\langle\Phi \rangle = v_{24} {\rm diag} (2, \ 2, \ 2, \ -3, \ -3)$. 
And we also assume a Z$_3$ symmetry breaking term
\begin{equation}
W_{SB}=  M^{SB}_i \, { \bf \bar{5}_{L(+)i}} 5'_{L(0)i} , 
\end{equation}
which induces  $\bar{5}_L \leftrightarrow \bar{5}'_L$ mixing
as follows:
\begin{eqnarray}
 \overline{5}'_{L(0)i}& =& 
c_i  \overline{5}^{q\ell}_{Li} +  
s_i  \overline{5}^{heavy}_{Li}   , \nonumber \\[4pt]
 \overline{5}_{L(+)i}& =& -s_i  \overline{5}^{q\ell}_{Li} 
 +  c_i  \overline{5}^{heavy}_{Li}   ,
\end{eqnarray}
where 
\begin{eqnarray}
s_i^{(a)} &=& \frac{ M^{(a)} }{ \sqrt{ (M^{(a)})^2 +
(M_i^{SB})^2} },\nonumber \\[4pt] 
  c_i^{(a)}& =& \frac{ M_i^{SB} }{ 
\sqrt{ (M^{(a)})^2 +(M_i^{SB})^2} },
\end{eqnarray}
$M^{(2)} = M_5 +3 g_5 v_{24}$, and $M^{(3)} = M_5 -2 g_5 v_{24}$.
Therefore, we obtain the following effective $R$-parity violation-terms:
\begin{equation}
W_{\not\!R}^{eff} = {  s_i^{(a)} s_j^{(b)}} \lambda_{ijk} 
\overline{5}^{ql(a)}_{Li}\overline{5}^{ql(b)}_{Lj} 10_{Lk}  ,
\end{equation}
Hereafter, for simplicity, we denote $\overline{5}^{ql(a)}_{Li}$ as
$\overline{5}^{(a)}_{Li}$.
 
We take the parameter values as follows:
\begin{eqnarray}
&&{} M^{(2)} \sim M_{GUT}, \ \ M^{(3)} \sim m_{SUSY},  \nonumber \\[4pt]
&&{} M_i^{SB} \sim M_{GUT} \times 10^{-1},
\end{eqnarray}
so that we obtain  values of the mixing parameters 
$s_i^{(2)} \simeq 1$ and, $c_i^{(2)} \simeq { M_i^{SB} }/{ M^{(2)} }
\sim  10^{-1}$ for the doublet components, and $ s_i^{(3)} \simeq 
{ M^{(3)} }/{M_i^{SB} } \sim 10^{-12}$ and $c_i^{(3)} \simeq 1$ 
for the triplet components. 
Since the unwelcome $R$-parity violation-terms 
$d^c_R d^c_R u^c_R$ and $d^c_R (e_L u_L -\nu_L d_L)$ are
suppressed by the factors $s^{(3)} s^{(3)} \sim 10^{-24}$ and
$s^{(3)} s^{(2)} \sim 10^{-12}$, respectively,
the proton decay due to the $R$-parity violation-terms is
suppressed by the factor of $10^{-36}$.
On the other hand, the $R$-parity violation-terms
$(e_L \nu_L -\nu_L e_L) e^c_R$ are of the order of 
$s^{(2)} s^{(2)} \sim 1$.

Note that $ M_d \neq M_e^T$ in the present model, because
\begin{equation}
M_d^\dagger= C^{(3)} Y_d v_d \ , \ \ \ 
M_e^* = C^{(2)} Y_d v_d , 
\end{equation}
where
$C^{(3)} = {\bf 1} +O(10^{-24})$
and $C^{(2)} \sim 10^{-1}$.

\section{Neutrino mass matrix}

First, we calculate a radiative mass from the diagram Fig.1:
\begin{eqnarray}
(M_{rad})_{ij} & \propto &
s_i s_j s_k s_n \lambda^*_{ikm} \lambda^*_{jnl}
  (M_e)^*_{kl} (\widetilde{M}_{eLR}^{2T})^*_{mn}\nonumber \\[4pt]
&&{} + 
(i \leftrightarrow j)   ,
\end{eqnarray}
where $s_i=s_i^{(2)}$.

\begin{figure}[hb]
\unitlength=0.7cm
\begin{flushright}
\begin{picture}(9,3.6)
\thicklines
%
%
\put(0.5,1){\line(1,0){2}}
\put(1.5,1){\vector(1,0){0}}
\put(0.8,1.3){$\nu_j$}
\put(2.5,1){\circle*{0.2}}
\multiput(2.5,1)(0.5,0){8}{\line(1,0){0.3}}
\put(3.5,1){\vector(1,0){0}}
\put(3,0.3){$\tilde{e}_R$}
\put(4.5,1){\circle*{0.2}}
\put(5.5,1){\vector(1,0){0}}
\put(5.5,0.3){$\tilde{e}_L$}
\put(6.5,1){\circle*{0.2}}
\put(4.2,-0.3){$\widetilde{M}^{2}_{eLR}$}
\put(6.5,1){\line(1,0){2}}
\put(8,1){\vector(-1,0){0}}
\put(8,1.3){$\nu_i^c$}
\put(4.5,1){
\qbezier(-2,0)(-2.01,0.35)(-1.88,0.68)
\qbezier(-1.88,0.68)(-1.78,1.03)(-1.53,1.29)
\qbezier(-1.53,1.29)(-1.3,1.55)(-1,1.73)
\qbezier(-1,1.73)(-0.69,1.9)(-0.35,1.97)
\qbezier(-0.35,1.97)(0,2.03)(0.35,1.97)
\qbezier(2,0)(2.01,0.35)(1.88,0.68)
\qbezier(1.88,0.68)(1.78,1.03)(1.53,1.29)
\qbezier(1.53,1.29)(1.3,1.55)(1,1.73)
\qbezier(1,1.73)(0.69,1.9)(0.35,1.97)
}
\put(4.4,3){\circle*{0.2}}
\put(4.2,3.3){$M_e$}
\put(2.3,2.3){$e_L$}
\put(6.6,2.3){$e_R$}
%
%
%
%
\end{picture}
\caption{Radiative generation of neutrino Majorana mass}
\label{fig:numass}
\end{flushright}
\end{figure}
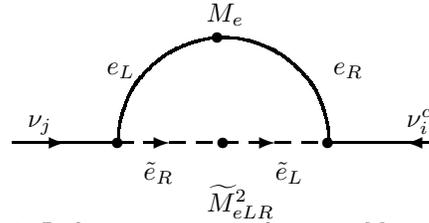

When we define
\begin{equation}
 K =  (S M_e L^T)^* \ ,
\end{equation}
\begin{equation}
\lambda_{ijk} = \varepsilon_{ijl} L_{lk} \ ,
\end{equation}
\begin{equation}
S={\rm diag}(s_1^{(2)}, s_2^{(2)}, s_3^{(2)}) \simeq {\bf 1} \ ,
\end{equation}
we can express $M_{rad}$ as
\begin{equation}
(M_{rad})_{ij} = m_0^{-1} s_i s_j \varepsilon_{ikm} 
\varepsilon_{jln} K_{ml} K_{nk},
\end{equation}
where
\begin{equation}
m_0^{-1} = \frac{2}{16 \pi^2} (A+\mu^{(2)}\tan\beta)
\frac{ \ln({m_{\tilde{e}_R}^2}/{m_{\tilde{e}_L}^2})}{
m_{\tilde{e}_R}^2-m_{\tilde{e}_L}^2} .
\end{equation}

Next, we calculate contributions from
the non-vanishing sneutrino VEV $\langle \widetilde{\nu} \rangle \neq 0$.
In the present model, the VEV of sneutrino is exactly zero at tree level,
because of the Z$_3$ symmetry.
However, only an effective $m_{HLi}^2$-term can appear 
via the loop diagram
$\overline{H}_d \rightarrow (\overline{5}^{ql}_L)^c + (10_L)^c 
\rightarrow \overline{5}_L^{ql}$ (Fig.~2), 
which gives
\begin{equation}
(m^2_{HL_i})_{eff} \propto s_i s_j \lambda_{ijk} (M_e)_{jk}
= s_i \varepsilon_{ijk} K^*_{jk} \ .
\end{equation}
Since $\langle \widetilde{\nu}_i \rangle \propto (m^2_{HL_i})^*_{eff}$,
we obtain
\begin{equation}
(M_{VEV})_{ij} =  \xi m_0^{-1} s_i s_j \varepsilon_{ikl} 
\varepsilon_{jmn} K_{kl} K_{mn},
\end{equation}
where $\xi$ is a relative ratio of $M_{VEV}$ to $M_{rad}$.

\begin{figure}[hb]
\unitlength=0.7cm
\begin{center}
\begin{picture}(9,4)
\thicklines
%
%
\put(0.5,1){\line(1,0){0.2}}
\put(0.8,1){\line(1,0){0.2}}
\put(1.1,1){\line(1,0){0.2}}
\put(1.4,1){\line(1,0){0.2}}
\put(1.1,1){\line(1,0){0.2}}
\put(1.8,1){\line(1,0){0.2}}
\put(1.1,1.3){$\overline{5}^{ql}_{Li}$}
\put(2.1,1){\line(1,0){0.2}}
\put(2.4,1){\line(1,0){0.2}}
\put(2.5,1){\circle*{0.2}}
\put(2.5,1){\line(1,0){4}}
\put(4.2,1.3){$(\overline{5}^{ql}_{Lj})^c$}
\put(6.5,1){\circle*{0.2}}
\put(6.5,1){\line(1,0){0.2}}
\put(6.8,1){\line(1,0){0.2}}
\put(7.1,1){\line(1,0){0.2}}
\put(7.4,1){\line(1,0){0.2}}
\put(7.4,1.3){$\overline{H}_d$}
\put(7.7,1){\line(1,0){0.2}}
\put(8.1,1){\line(1,0){0.2}}
\put(8.4,1){\line(1,0){0.2}}

\put(4.5,1){
\qbezier(-2,0)(-2.01,0.35)(-1.88,0.68)
\qbezier(-1.88,0.68)(-1.78,1.03)(-1.53,1.29)
\qbezier(-1.53,1.29)(-1.3,1.55)(-1,1.73)
\qbezier(-1,1.73)(-0.69,1.9)(-0.35,1.97)
\qbezier(-0.35,1.97)(0,2.03)(0.35,1.97)
\qbezier(2,0)(2.01,0.35)(1.88,0.68)
\qbezier(1.88,0.68)(1.78,1.03)(1.53,1.29)
\qbezier(1.53,1.29)(1.3,1.55)(1,1.73)
\qbezier(1,1.73)(0.69,1.9)(0.35,1.97)
}
\put(4.2,3.3){$(10_{Lk})^c$}
\put(2.2,0.3){$s_i \lambda_{ijk}$}
\put(6.0,0.3){$c_j(Y_d^*)_{jk}$}
%
%
%
%
%
%
\end{picture}
\caption{Effective $\overline{5}^{ql}_L \overline{H}_d^\dagger$ term}
\end{center}
\end{figure}
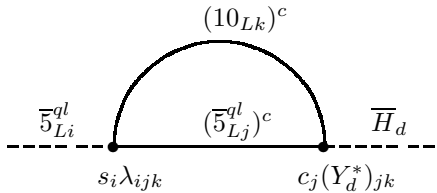

In conclusion, we obtain the following general form of 
the neutrino mass matrix\cite{SUSY-Koide2}
\begin{eqnarray}
(M_\nu)_{ij}& =&   m_0^{-1} s_i s_j \varepsilon_{ikl} 
\varepsilon_{jmn}\nonumber \\[4pt]
&&{} \left(   K_{kn} K_{ml} 
+ \xi  K_{kl} K_{mn} \right),
\end{eqnarray}
i.e.
\begin{equation}
M_\nu =   m_0^{-1} S \left[ A (1+\xi) +B \right] S
\end{equation}
where
\begin{eqnarray}
A&= & (K -K^T)(K- K^T) - {\bf 1} {\rm Tr}(KK-K K^T) , \nonumber \\[4pt]
B& = & \left( K+K^T -{\bf 1} {\rm Tr}K \right) {\rm Tr}K \nonumber \\[4pt]
&&{}  -(KK+K^TK^T)+ {\bf 1} {\rm Tr}(KK)   ,
\end{eqnarray}
Note that $A$ is a rank-1 matrix which is independent of
the diagonal elements of $K$, $K_{11}$, $K_{22}$ and $K_{33}$.

\section{A simple example}

Hereafter, we discuss the quantities on the flavor basis where
$M_e$ is diagonal.

Let us consider a simple form of $K$ which gives $A \gg B$.
We assume the following form of ${K}$: 
\begin{equation}
 {K}/m_{0K} =
\left(
\begin{array}{ccc}
1 & 1 & 1 \\
0 & 0 & 0 \\
0 & 0 & 0 
\end{array} \right) \ + \varepsilon {\bf 1} ,
\end{equation}
where $m_{0K}$ is a constant with a dimension of mass.
The form (21) means that in the limit of $\varepsilon =0$,
the coefficients $ {\lambda}_{ijk}$ of the $R$-parity violation terms 
are given by $ {\lambda}_{ij1}= {\rm const} \equiv \lambda$ 
and ${\lambda}_{ij2}= {\lambda}_{ij3}=0$,
i.e.
 $\bar{5}_{Li} \bar{5}_{Lj}$ (i.e. 
$\ell_{Li} \ell_{Lj}$) can couple only to $10_{L1}$ (i.e. $e_R^c$).
The assumption (21) leads to
\begin{equation}
 {M}_{\nu} = (1+\xi)A +B  ,
\end{equation}
where
\begin{equation}
 {A} = m_{0K}^2 
\left(
\begin{array}{ccc}
0 & 0 & 0 \\
0 & 1 & -1 \\
0 & -1 & 1 
\end{array} \right)  ,
\end{equation}
and
\begin{equation}
 {B} = m_{0K}^2 \varepsilon
\left(
\begin{array}{ccc}
2 \varepsilon  & 1 & 1 \\
1 & -(1+\varepsilon) & 0 \\
1 & 0 & -(1+ \varepsilon) 
\end{array} \right)  ,
\end{equation}
$m^\nu_0 = m_0^{-1} m_{0K}^2$ and we have put $S={\bf 1}$.
The mass matrix (22) gives the following 
eigenvalues and mixings:
\begin{eqnarray}
m_{\nu 1}& =& (\sqrt{3}-1-2 \varepsilon) \varepsilon 
m^\nu_0  \ , \nonumber \\[4pt]
m_{\nu 2}& =& -(\sqrt{3}+1+2 \varepsilon) \varepsilon 
m^\nu_0\  , \\
m_{\nu 3}& =& 2(1+\xi -\varepsilon - \varepsilon^2) m^\nu_0\  , 
\nonumber 
\end{eqnarray}
\begin{equation}
 {U}_{\nu} = 
\left(
\begin{array}{ccc}
\sqrt{ \frac{\sqrt{3}+1}{2\sqrt{3}} } & 
-\sqrt {\frac{\sqrt{3}-1}{2\sqrt{3}} } & 0 \\
\frac{1}{\sqrt{6+2\sqrt{3}}} & 
\frac{1}{\sqrt{6-2\sqrt{3}}} & -\frac{1}{\sqrt2} \\
\frac{1}{\sqrt{6+2\sqrt{3}}} & 
\frac{1}{\sqrt{6-2\sqrt{3}}} & \frac{1}{\sqrt2} 
\end{array} \right) .
\end{equation}
Note that 
the structure of the mixing matrix $ {U}_\nu$, (26), 
is independent of the parameters $\xi$ and $\varepsilon$.
Therefore, we obtain the neutrino mixing parameters
\begin{equation}
\tan^2 \theta_{solar}=\frac{\sqrt{3}-1}{\sqrt{3}+1}= 0.268,
\end{equation}
together with $\sin^2 2 \theta_{atm} = 1$ and $|U_{13}|^2 = 0$,
and the ratio of the neutrino mass squared
\begin{equation}
R \equiv \frac{\Delta m^2_{21}}{\Delta m^2_{32}} 
\simeq  \frac{\sqrt{3}(1+2{  \varepsilon})}{(1+\xi)
(1+\xi -2{  \varepsilon})}
{  \varepsilon^2} .
\end{equation}
Roughly speaking, the these results are favorable to the 
recent neutrino data \cite{McGrew,Wang}.
Although the predicted value of $\tan^2 \theta_{solar}$ is
somewhat smaller than the observed best fit value, the 
value can suitably be adjusted by a small deviation of $ {S}$ from
$ {S} = {\bf 1}$ and the renormalization group equation effects.

\section{Conclusions}

Based on a SUSY SU(5) GUT model with harmless $R$-parity violation,
we have proposed a neutrino mass matrix with a simple form, which are 
given by sum of the radiative masses plus nonvanishing sneutrino
VEV contributions.
The model with a simple assumption (21) leads to reasonable results
\begin{eqnarray}
\sin^2 2\theta_{23}=1, \ \ |U_{13}|=0, \nonumber \\[4pt]
\tan^2 \theta_{12}=\frac{\sqrt{3}-1}{\sqrt{3}+1}= 0.268,  
\end{eqnarray}
independently of the parameters $\varepsilon$ and $\xi$.
However, at present, it is an open question why we should choose 
such the simple form of $K$.

Although the form (21) is only an example, we can, at least, say
that, as the origin of the neutrino masses,  we should seriously 
take a possibility of the radiative mass generation mechanism 
as well as that of the seesaw mechanism.


\begin{figure}
\epsfxsize170pt 
\figurebox{180pt}{170pt}{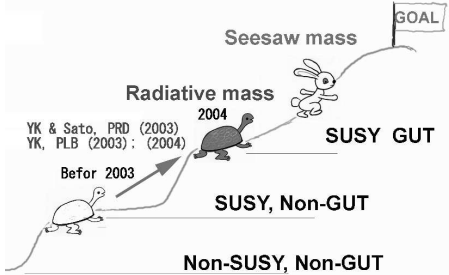} 
\caption{The status of the origin of the neutrino mass in 2004:
the radiative neutrino mass hypothesis has considerably become 
plausible than before}
\label{fig:radk}
\end{figure}

\section*{Acknowledgments}
This work was supported by the Grant-in-Aid for Scientific
Research, the Ministry of Education, Science and Culture
(Grant Number 15540283).



\end{document}